\documentclass{article}
\usepackage{url}
\usepackage{graphicx}
\usepackage{xcolor}
\usepackage{float}
\usepackage[acronym]{glossaries}
\usepackage{glossaries-extra}
\usepackage{subfig}
\usepackage{amsmath,amssymb,amsfonts}
\usepackage{xspace}
\usepackage{pifont}
\usepackage[%
  bibstyle=ieee,
  citestyle=numeric-comp,
  isbn=true,
  doi=false,
  sorting=none,
  url=true,
  defernumbers=true,
  bibencoding=utf8,
  backend=biber,
  maxnames=2,
  minnames=1
]{biblatex} 
\GlsXtrEnableEntryCounting{acronym}{2}

\setabbreviationstyle[acronym]{long-short}

\newcommand{\zh}{$ZH,H\rightarrow \tau\tau$\xspace}
\newcommand{\z}{$Z/\gamma^{*}\rightarrow\tau\tau$\xspace}
\newcommand{\qq}{$Z/\gamma^{*}\rightarrow qq$\xspace}
\newcommand{\ptmomentum}{\ensuremath{p_{\mathrm{T}}}\xspace}
\newcommand{\ptvistrue}{\ensuremath{p_{\mathrm{T}}^{\mathrm{vis,true}}}\xspace}
\newcommand{\ptvispred}{\ensuremath{p_{\mathrm{T}}^{\mathrm{vis,pred}}}\xspace}
\newcommand{\ptjet}{\ensuremath{p_{\mathrm{T}}^{\mathrm{jet}}}\xspace}
\newcommand{\ptgenjet}{\ensuremath{p_{\mathrm{T}}^{\mathrm{gen-jet}}}\xspace}

\newacronym{hps}{HPS}{hadron-plus-strips}
\newacronym{cms}{CMS}{Compact Muon Solenoid}
\newacronym{lhc}{LHC}{large hadron collider}
\newacronym{bdt}{BDT}{boosted decision tree}
\newacronym[longplural={boosted regression trees}, shortplural=BRTs]{brt}{BRT}{boosted regression tree}
\newacronym{dl}{DL}{deep learning}
\newacronym{iqr}{IQR}{interquartile range}
\newacronym{ecal}{ECAL}{electromagnetic calorimeter}
\newacronym[longplural={decay modes}, shortplural=DMs]{dm}{DM}{decay mode}
\newacronym[longplural={recurrent neural networks}, shortplural=RNNs]{rnn}{RNN}{recurrent neural networks}
\newacronym{ild}{ILD}{International Large Detector}
\newacronym{idea}{IDEA}{Innovative Detector for Electron-positron Accelerator}
\newacronym{atlas}{ATLAS}{A Toroidal LHC Apparatus}
\newacronym{tpf}{TPF}{Tau Particle Flow}
\newacronym{clic}{CLIC}{Compact Linear Collider}
\newacronym{sota}{SOTA}{state-of-the-art}
\newacronym{cld}{CLD}{CLIC Like Detector}
\newacronym{sm}{SM}{Standard Model of particle physics}
\newacronym{bsm}{BSM}{physics beyond the SM}
\newacronym{nn}{NN}{neural network}
\newacronym[longplural={branching ratios}, shortplural=BRs]{br}{BR}{branching ratio}
\newacronym{pp}{pp}{proton-proton}
\newacronym{ee}{$e^+e^-$}{electron-positron}
\newacronym[longplural={field programmable gate arrays}, shortplural=FPGAs]{fpga}{FPGA}{field programmable gate array}
\newacronym{clicdet}{CLICdet}{CLIC like detector}
\bibliography{main.bib}

\title{A unified machine learning approach for reconstructing hadronically decaying tau leptons}
\author{
  Tani, Laurits$^1$\\ \texttt{laurits.tani@cern.ch} \and
  Seeba, Nalong-Norman$^1$ \\ \texttt{nalong-norman.seeba@cern.ch} \and 
  Vanaveski, Hardi$^{1,2}$ \\ \texttt{hvanav@taltech.ee} \and 
  Pata, Joosep$^1$ \\ \texttt{joosep.pata@cern.ch} \and
  Lange, Torben$^1$ \\ \texttt{torben.lange@cern.ch} \and 
}
\date{%
    {\footnotesize $^1$\emph{National Institute Of Chemical Physics And Biophysics (NICPB), Rävala pst. 10, 10143 Tallinn, Estonia}\\
    $^2$\emph{Tallinn University of Technology (TalTech), Ehitajate tee 5, 19086 Tallinn, Estonia}\\}
}

\begin{document}

\maketitle

\begin{abstract}
    Tau leptons serve as an important tool for studying the production of Higgs and electroweak bosons, both within and beyond the Standard Model of particle physics.
    Accurate reconstruction and identification of hadronically decaying tau leptons is a crucial task for current and future high energy physics experiments.
    Given the advances in jet tagging, we demonstrate how tau lepton reconstruction can be decomposed into tau identification, kinematic reconstruction, and decay mode classification in a multi-task machine learning setup.
    Based on an electron-positron collision dataset with full detector simulation and reconstruction, we show that common jet tagging architectures can be effectively used for these sub-tasks.
    We achieve comparable momentum resolutions of 2--3\% with all the tested models, while the precision of reconstructing individual decay modes is between 80--95\%.
    We find ParticleTransformer to be the best-performing approach, significantly outperforming the heuristic baseline.
    This paper also serves as an introduction to a new publicly available $\mathtt{Fu}\tau\mathtt{ure}$ dataset for the development of tau reconstruction algorithms.
    This allows to further study the resilience of ML models to domain shifts and the efficient use of foundation models for such tasks.
\end{abstract}

\section{Introduction}
    %introduction tau in physics
    Tau leptons ($\tau$) serve as an important tool for tests of the \gls{sm} in the electroweak sector as well as for searches \gls{bsm} both at current and future high energy physics experiments.
    Due to its relatively high mass, the $\tau$ lepton couples strongly to the Higgs boson ($H$), enabling tests of Higgs boson coupling to third generation fermions~\cite{CMS:2022dwd,ATLAS:2022vkf}, and other Higgs boson properties like CP~\cite{ATLAS:2022akr, CMS:2021sdq}, rare processes like double Higgs~\cite{ATLAS:2022xzm,CMS:2022hgz} and precision measurements~\cite{ATLAS:2024wfv,CMS:2022kdi}.
    Furthermore, studying $\tau$ lepton properties --- such as its mass, lifetime and \glspl{br} --- makes it possible to test lepton universality of charged current between different fermion generations, while spin polarization measurements of the $\tau$ leptons permit us to probe neutral-current interactions~\cite{ALEPH:2001uca,BOYKO2001125,CMS:2023mgq,ALEPH:2005qgp,Dam:2021ibi,Pich:2020qna,ATLAS:2017xuc}.
    The $\tau$ can also be used to search for lepton flavor violating processes in $\tau$ lepton decays ~\cite{Dam:2021ibi,Pich:2020qna} as well as in decays of $Z$ bosons ~\cite{Dam:2018rfz,ATLAS:2021bdj} and Higgs bosons ~\cite{Harnik:2012pb,CMS:2021rsq,ATLAS:2023mvd} into a $\tau$ lepton and an electron or muon, forbidden in the \gls{sm} and a variety of other \gls{bsm} theories.
    
    %tau decay and properties / ML problem
    $\tau$ leptons have a very short lifetime of only $2.9\times10^{-13}$ seconds~\cite{ParticleDataGroup:2022pth}, short enough to decay before interacting with the detector material or before undergoing any radiative processes. In about a third of the cases, the $\tau$ decays into another, lighter lepton (electron or muon) and a neutrino.
    As the neutrino does not interact with the detectors used in high energy physics experiments, it can not be reconstructed and thus, leptonically decaying taus leave a signature in the detector that is mostly indistinguishable from an electron or muon for which there already exist dedicated reconstruction algorithms.
    
\begin{figure}[ht]
        \centering
        \includegraphics[width=0.8\textwidth]{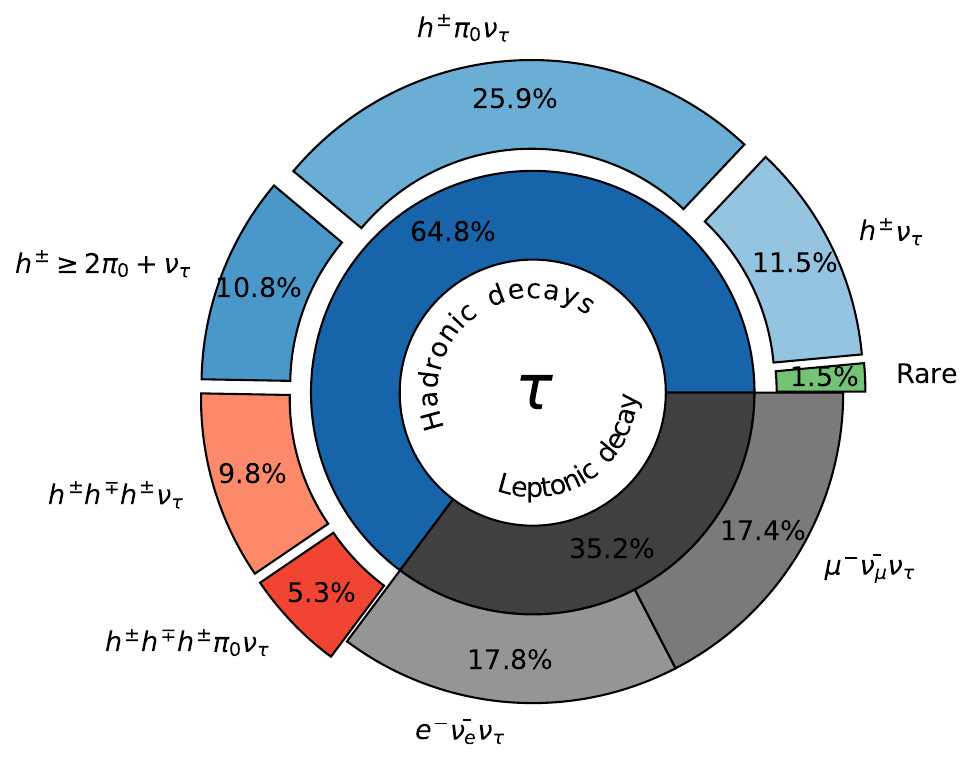}
        \caption{Overview of the branching fraction for various $\tau$ decay modes. The numbers are taken from the Particle Data Group~\cite{ParticleDataGroup:2022pth}.}
        \label{fig:tau_decays}
    \end{figure}

    More relevant for the context of this paper are hadronically decaying taus ($\tau_h$).
    In two thirds of the cases the $\tau$ lepton decays into a neutrino, an odd number of charged (typically pions $\pi^\pm$ or kaons $K^\pm$) hadrons ($h^\pm$) and a number of neutral pions ($\pi^0$).
    The latter particles decay almost instantly into photon pairs.
    Most dominant $\tau_h$ \glspl{dm} feature up to three $h^\pm$ and two or fewer $\pi^0$.
    An overview of the relative \glspl{br} for the different $\tau_h$ decays is given in Fig.~\ref{fig:tau_decays}.
    
    Correctly reconstructing the $\tau_h$ and identifying it from other particle signatures represents a significant combinatorial problem, which eluded a fully machine learning based solution for many years.
    Instead, in both $e^+e^-$ and \gls{pp} experiments, it relied on an accurate lifetime variable determination, $h^\pm$ and $h^0$ reconstruction, and their combination for reconstructing the tau candidates~\cite{Garcia-Abia:2001hmi,McNulty:2001jt,ALEPH:2001uca,Riles:1992rd}.
    In contemporary \gls{pp} experiments, a combination of heuristic algorithms, such as the \gls{hps}~\cite{cms2012performance,CMS:2015pac} is used for reconstruction, and machine learning-based methods for the $\tau_h$ identification.
    As a system of multiple final state particles, $\tau_h$ decays are on first glance especially difficult to differentiate from jets produced by other high energy processes such as the hadronization of a gluon or a quark.
    Thus, the problem of identifying a $\tau_h$~\cite{Russell:2827366,cms2012performance,ATLAS:2022aip,ATLAS:2019uhp,CMS:2022prd,CMS:2015pac,Huang:2023ssr, Madysa:2015gxj,atlas2017measurement} lies in the realm of "jet-tagging", which in recent years has been thoroughly explored with machine learning techniques, usually focusing on the differentiation of heavy quark jets, such as b-- and c--quarks, from lighter (u, d, s) quark and gluon jets with examples given in Refs.~\cite{Mondal:2024nsa,Dreyer:2020brq,Cheng:2017rdo}.
    
    %previous work and new problems (decay mode/regression)
    Substantial advances in jet-tagging have been made by exploiting deep learning techniques, such as transformers, originally developed for language modeling.
    As demonstrated previously in \cite{Lange:2023gbe}, these techniques can also be used to effectively identify $\tau_h$, with the two tested architectures, LorentzNet~\cite{Gong:2022lye} and ParticleTransformer~\cite{Qu:2022mxj}, outperforming more typical, heavily optimized approaches such as \gls{hps} + DeepTau~\cite{CMS:2022prd} even without any fine-tuning of aforementioned models by domain experts.
    Here, we show that such models can also be used for reconstructing the properties of the $\tau_h$.

    In addition to the correct identification of the $\tau_h$ candidates from jets, existing $\tau_h$ reconstruction chains developed for \gls{pp} experiments, such as ATLAS and CMS, aim to determine the exact \gls{dm} of the $\tau_h$ decay and to precisely reconstruct the $\tau_h$ momentum. 
    
    %state of the art

    Only the $\tau_h$ \glspl{dm} with the largest \gls{br} are targeted by the state-of-the-art algorithms.
    These \glspl{dm} include $h^\pm$; $h^\pm + \pi^0$; $h^\pm + \geq 2\pi^0$; $h^\pm h^\mp h^\pm$ and $h^\pm h^\mp h^\pm + \geq \pi^0$, with the rest of the $\tau_h$ decays being usually classified into the ``Rare'' or ``Other'' category.
    Similarly to \gls{ee} experiments, where the \glspl{dm} are usually classified by \glspl{bdt}~\cite{xu2017detectors} or neural networks~\cite{giagu2022tau}, the classification in \gls{pp} experiments is done either with a combinatorial approach~\cite{chen2022tau,CMS:2015pac} or using \glspl{bdt} on top of reconstructed $\tau_h$ candidates~\cite{ATLAS:2015boj}. 
    The classification precision for $h^\pm$ and $h^\pm h^\mp h^\pm$ is $\geq 90\%$ for both \gls{pp} as well as \gls{ee} collider experiments. However, as reconstructing multiple $\pi^0$s in \gls{pp} collider experiments is a more challenging task, with the precision for such \glspl{dm} featuring $\pi^0$s being $<60\%$~\cite{CMS:2015pac,chen2022tau,ATLAS:2015boj,Saxton:2634314}
    The corresponding precision for \gls{ee} collider experiments is $\geq 85\%$~\cite{giagu2022tau,xu2017detectors,Tran:2015nxa}. 
    
    In \gls{pp} experiments, the $\tau_h$ kinematic reconstruction is done both in a combinatorial approach~\cite{CMS:2015pac} and a mixture of combinatorics and boosted regression trees~\cite{ATLAS:2015boj,hubner2018measurement}.
    The performance of such approaches is comparable, achieving an average energy resolution of $\sim$14\%~\cite{CMS:2015pac} for the energies ranging from 30 GeV to 300 GeV, with the resolution degrading at higher energies for up to 20\% for higher energies. % This is about CMS
    The energy resolution is better in the core region of the jet, being in the order of 5--7\%, while the tail resolution is 18--30\%~\cite{ATLAS:2015boj, hubner2018measurement}. % This is about ATLAS
    Despite the $\tau_h$ kinematic reconstruction usually being done with respect to $\tau_h$ energy, then, similarly to jet reconstruction, we are regressing the transverse momentum ($p_T$) of the $\tau_h$.

    In addition to the above mentioned tasks of $\tau_h$ identification, kinematic regression and its decay mode reconstruction, ML methods have also been studied for various other scenarios in $\tau$ physics.
    These include, for example, mass reconstruction of heavy gauge boson decaying into $\tau$ leptons~\cite{Krishnan:2023fev} and boosted di-$\tau$ system identification~\cite{Tamir:2023aiz}.

    % this work
    In this work, we aim to expand the applications of deep learning based jet tagging algorithms to the problems of $\tau_h$ energy regression and decay mode classification, with the ultimate goal of providing a recipe for a complete $\tau_h$ reconstruction in an end-to-end approach for a given detector configuration.
    As both kinematic and decay mode reconstruction would only be run on high-purity jets that are identified as $\tau$-jets, background samples are not used in these studies.
    Furthermore, the assessment of the mis-identification rate of electrons and muons as $\tau_h \rightarrow \pi^\pm + \nu_\tau$ are left for future studies.
    To further facilitate the development and studies of new $\tau_h$ reconstruction algorithms, we also provide a well documented version of our $\mathtt{Fu}\tau\mathtt{ure}$ dataset.
    
    The dataset and an overview of its features are given in Sec.~\ref{sec:dataset} with further details along with the dataset itself in Ref.~\cite{dataset}.
    We formulate $\tau_h$ reconstruction as a set of machine learning tasks in Sec.~\ref{sec:reconstruction} and demonstrate how three different types of models with a varying degree of expressiveness and priors can be employed for the tasks.
    We study the performance of the models on the tasks in Sec.~\ref{sec:results} and give a short summary and outlook to future work in Sec.~\ref{sec:outlook}.

\section{The $\mathtt{Fu}\tau\mathtt{ure}$ dataset}\label{sec:dataset}
    With this paper, we provide the first version of the $\mathtt{Fu}\tau\mathtt{ure}$ dataset.
    This dataset includes $\sqrt{s}=380$ GeV $e^+e^-$ Monte Carlo samples with \z, \zh and \qq processes with approximately 2 million events for each process.
    For the simulation and reconstruction we used the \gls{clic} detector~\cite{Linssen:2012hp} setup.
    The dataset is updated with respect to~\cite{Lange:2023gbe} with more simulated events, and is now released with documentation for the first time.
    The generation of the events is described in Sec.~\ref{sec:mc-generation} with an overview and description of the features available in Sec.~\ref{sec:features}.
    The public dataset along with more technical information can be found in Ref.~\cite{dataset}.
    For the studies in this paper, only $\tau_h$ jets in the \z and \zh datasets were used.

    \subsection{Monte Carlo samples and event reconstruction}\label{sec:mc-generation}
        The $\mathtt{Fu}\tau\mathtt{ure}$ dataset is generated using Pythia8~\cite{10.21468/SciPostPhysCodeb.8} with the same generator settings as in Ref.~\cite{Lange:2023gbe,Amhis:2021cfy}.
        After generation the events undergo a full detector simulation using Geant4~\cite{Agostinelli:2002hh} with the \gls{clicdet} (CLIC\_o3\_v14)~\cite{CLICdp:2017vju} setup before being reconstructed using the Marlin reconstruction code~\cite{Gaede:2006pj} and the \texttt{Key4HEP}~\cite{Ganis:2021vgv} software and prepared in the \texttt{EDM4HEP}~\cite{Gaede:2022leb} format.
        The features in our dataset are then extracted using the particle flow candidates found by processing the reconstructed events with the PandoraPF algorithm~\cite{Marshall:2012ry,Marshall:2015rfa}. 
    
        The \gls{clicdet} design has been thoroughly studied over the recent years, and its design is similar to the \gls{cld}, foreseen for the upcoming Future Circular Collider (FCC), thus providing a relevant benchmark scenario for physics at future $e^+e^-$ experiments.
        Similarly to other contemporary detectors at hadron colliders, the \gls{clicdet} detector features a layered design with a high precision tracking system featuring a silicon pixel and tracking detector, a Si-W electromagnetic sampling calorimeter and and a steel hadronic sampling calorimeter encased in a 4T solenoid and a dedicated muon system.
        The expected physics performance with some preliminary studies is discussed in Ref.~\cite{Abramowicz:2016zbo}.
        
        A potentially important background for the $\tau_h$ reconstruction from the overlay of $\gamma\gamma\rightarrow\,$hadrons is currently not included in our simulation. The study and inclusion of this background is left for future iterations of this dataset.
    
    \subsection{Input features and validation}\label{sec:features}
        The basis of our dataset are particle flow candidates from PandoraPF with four momenta, charge, and candidate labels for electrons ($e$), muons ($\mu$), photons ($\gamma$), charged hadrons ($h^\pm$), and neutral hadrons ($h^0$).
        These candidates are clustered into jets using the generalized $k_T$ algorithm for $e^+e^-$ collisions (\texttt{ee\_genkt})~\cite{Boronat:2016tgd} with parameters of $p=-1$, $R=0.4$, and a minimum jet transverse momentum of $\ptmomentum \geq 5~\mathrm{GeV}$ to serve as the seeds for $\tau_h$ identification.
        The dataset contains the four momenta of these reconstructed jets, and the four momenta, charge and the particle label of the PandoraPF candidates within them.
        
        As $\tau$-leptons have a small but finite lifetime, corresponding to a travel distance of a few mm in the detector, variables sensitive to this special topology such as transverse ($d_{xy}$) and longitudinal ($d_z$) impact parameters of the tracks of charged particles have long been used for $\tau_h$ identification. 
        We use the linear approximations to define impact parameters as in Ref.~\cite{Lange:2023gbe}, with more details to be found in Refs.~\cite{Kuhr:1998jk,Kramer:2006zz}. However, the full evaluation of the effects of impact parameters on the tasks is left for a future study.
        
        The ground truth is based on stable particles at the generator level, before detector simulation.
        These particles are clustered into generator-level jets with the same algorithm as used for the clustering of the reconstructed jets, but without a cut on generator-level jet transverse momentum, \ptgenjet.
        The generator-level jets are matched to generator-level $\tau_h$ as well as reconstructed jets within the angular distance of $\Delta R < 0.3$.
        For each reconstructed jet, we can then define up to three target values related to $\tau_h$ reconstruction:
        \begin{itemize}
            \item a binary flag \texttt{isTau} if it was matched to a $\tau_h$;
            \item if matched, the categorical decay mode of the $\tau_h$ in terms of the number of charged and neutral hadrons $\mathrm{DM}^{\mathrm{true}} \in \{0,1,\dots,15\}$;
            \item if matched, the visible (i.e. neutrinoless and reconstructable) transverse momentum \ptvistrue of the generator-level $\tau_h$.
        \end{itemize}
        
        Thus, the dataset consists of reconstructed jets, with a variable number of reconstructed particles per jet, and with up to three target labels for each reconstructed jet.
        While the models we subsequently study are invariant to particle ordering within the jet, we sort the particles in each jet in \ptmomentum descending order in the interest of clarity.
        
\begin{figure}[ht]
            \centering
            \includegraphics[width=0.49\textwidth]{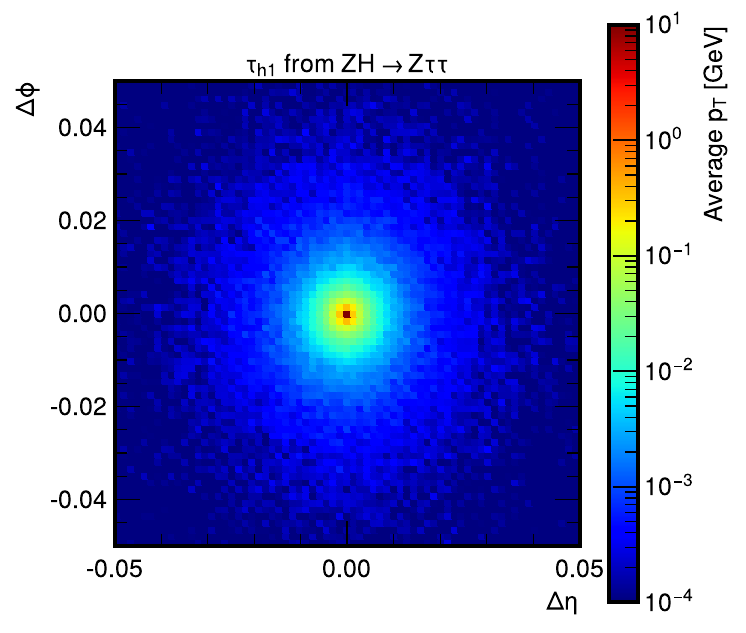}
            \includegraphics[width=0.49\textwidth]{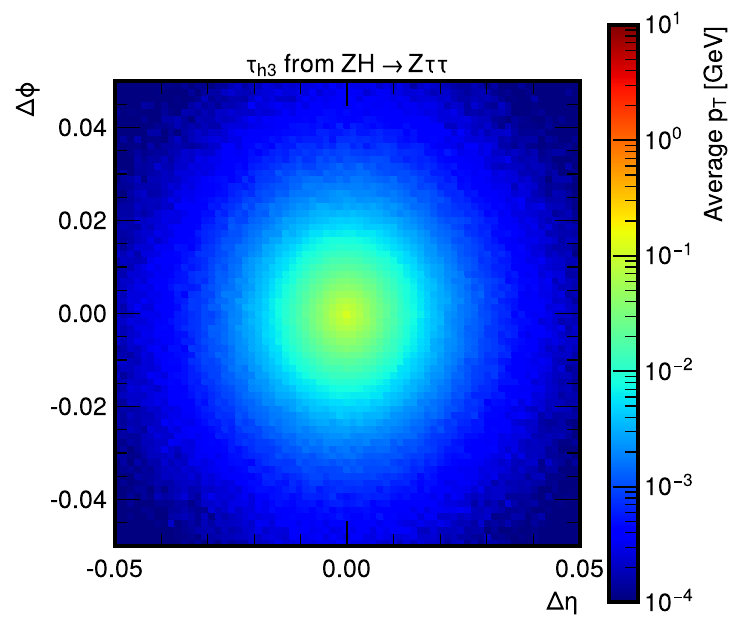}
            \caption{Average \ptmomentum of particle flow candidates per bin in the $\Delta\eta-\Delta\phi$ plane around the core of the jet, aggregated for jets, matched to a $\tau_{h_1}$ (\textbf{Left}), and $\tau_{h_3}$ (\textbf{Right}) $\tau_h$ decay modes across the \zh sample.}
            \label{fig:jetstructurevsdecay}
        \end{figure}

\begin{figure}[ht]
            \centering
            \includegraphics[width=0.49\textwidth]{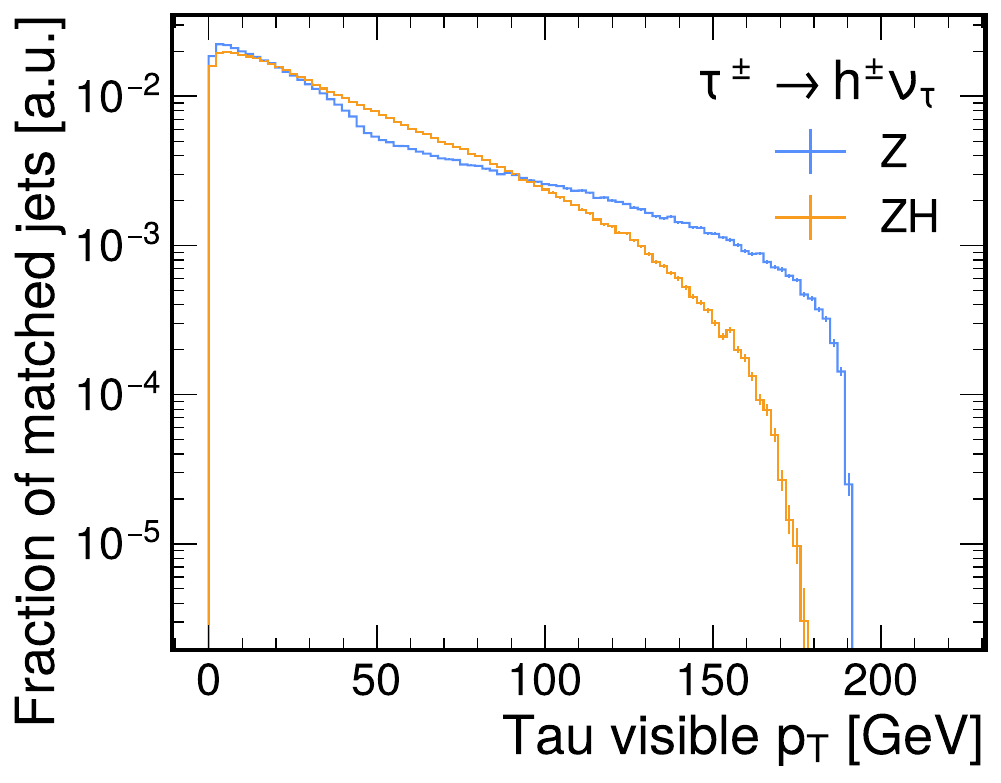}
            \includegraphics[width=0.46\textwidth]{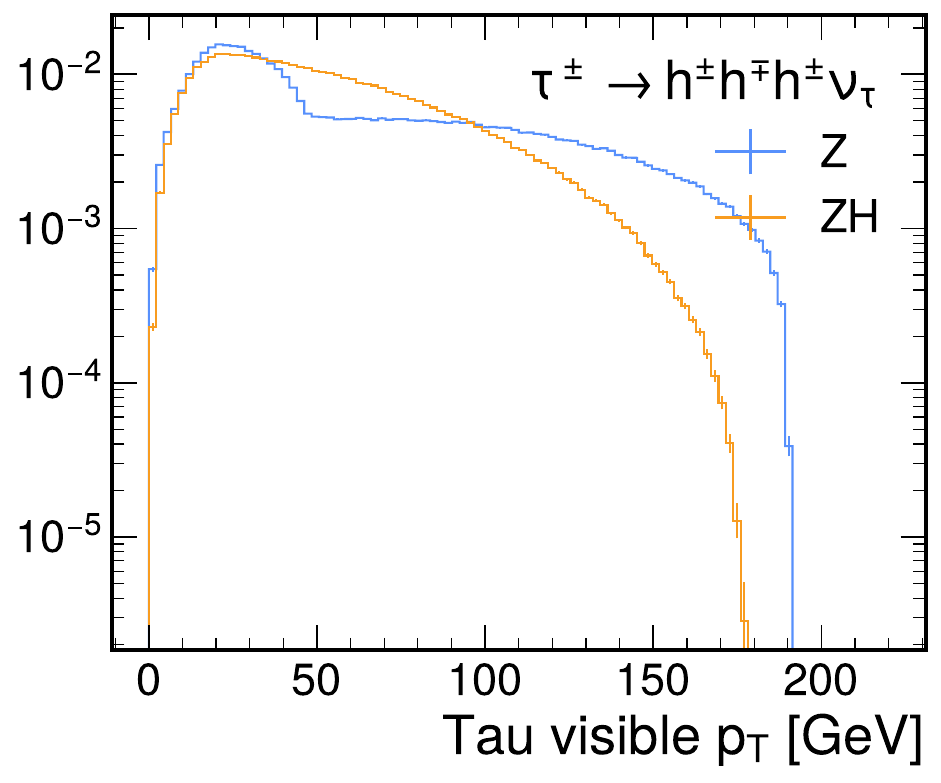}
            \caption{Distribution of transverse momentum of the visible $\tau_h$ component in the \z and \zh datasets for $\tau_{h1}$ (\textbf{Left}) and $\tau_{h3}$ (\textbf{Right}).}
            \label{fig:tau_visgenenergy}
        \end{figure}

        The dataset has been extensively investigated and tested for consistency and can be validated with the provided software found in Ref.~\cite{software} and Ref.~\cite{dataset}.
        As an example, Figs.~\ref{fig:jetstructurevsdecay} and \ref{fig:tau_visgenenergy} illustrate the two research problems at hand: the first figure shows the jet substructure for two different $\tau_h$ \glspl{dm} in the \zh sample --- decays into a single charged hadron ($\tau_{h_1}$) and three charged hadrons ($\tau_{h_3}$).
        The second figure shows the generator-level \ptmomentum of the visible $\tau_h$ decay products we regress from the jet constituents for $\tau_{h_1}$ and $\tau_{h_3}$.
        
\section{Tau reconstruction with ML}\label{sec:reconstruction}
    Reconstructing hadronically decaying tau leptons using ML can be defined as a multi-task machine learning problem:
    \begin{equation*}
        \Phi(\mathrm{jet\ features}, \mathrm{particle\ features}) \rightarrow \{\texttt{isTau}, \mathrm{DM}^{\mathrm{true}}, \ptvistrue\}\,\,,
    \end{equation*}
    where $\Phi$ is a trainable model.
    Note that $\Phi$ may consist of a single model, separate models, or even a single backbone model with fine-tuned output layers for each task.
    
    We addressed $\texttt{isTau}$ classification in Ref.~\cite{Lange:2023gbe}, finding that in this dataset, a transformer-based approach performs well compared to the alternatives for $\tau_h$ identification.
    
    In this paper, we address reconstructing the hadronic $\tau$ \glspl{dm}, as given in Fig.~\ref{fig:tau_decays}, from a $\tau_h$ candidate jet and regressing the visible transverse momentum of the generator-level $\tau_h$, \ptvistrue.
    
    We use the ParticleTransformer~\cite{qu2022particle} and LorentzNet~\cite{gong2022efficient} architectures as the main models due to their expressiveness on jet tasks.
    As a cross-check, we also test a simpler algorithm based on deep sets~\cite{zaheer2017deep} (DeepSet).
    While expressive, transformer-based models can be resource-intensive and challenging to run in real-time on constrained hardware such as \glspl{fpga}, or to include physics-informed priors.
    The performance of LorentzNet architecture indicates the usefulness of a strong inductive bias based on Lorentz symmetry.
    The DeepSet architecture serves as a simple cross-check model, which tests how much the previous, more expressive models add on top of a very simple baseline.
    Moreover, DeepSet-type models are straightforward to deploy on constrained hardware such as \glspl{fpga}, and may represent a practical trade-off where accuracy needs to be balanced with inference throughput and latency~\cite{Yaary:2023dvw, Motta:2024sqf}. 
    
    We thus use the $\mathtt{Fu}\tau\mathtt{ure}$ dataset to compare three model architectures on two different tasks.
    For each reconstructed PF candidate (see Sec.~\ref{sec:features}) in the jet, we use the same set of input features for all models and tasks, consisting of the particle kinematics $p_x, p_y, p_z, E$ and the additional features:
    \begin{itemize}
        \item particle charge, $q \in \{-1, 0, +1\}$;
        \item particle labels \texttt{isChargedHadron}, \texttt{isNeutralHadron},\texttt{isPhoton}, \texttt{isElectron}, \texttt{isMuon} based on particle flow reconstruction;
        \item $\log{\ptmomentum}$, $\log{E}$ of the particles;
        \item $\Delta \eta$, $\Delta \phi$ of the particle with respect to the jet;
        \item relative $\log\left({\ptmomentum / \ptjet}\right)$, $\log\left({E / E_{jet}}\right)$ with respect to the jet.
    \end{itemize}

    Currently no $\tau$ lifetime information was used for the momentum and decay mode reconstruction as their impact on the performance on a pure sample of $\tau_h$ is limited.
    However, the inclusion of these variables will more important in future studies with the inclusion of the $\gamma\gamma\rightarrow$ hadrons overlay background, which similarly to ``pileup'' from simultaneously occurring collisions at hadron colliders, could contaminate the jets with additional low energy particles, thus spoiling the momentum reconstruction.
    
    We pick the first 16 particles in \ptmomentum-descending order from each jet for the subsequent studies for efficient transformer training, which introduces a negligible fraction of lost particles.
    The ML models we study are invariant to particle ordering in the jet.
    
    For the training we use the \z sample, while the final result is evaluated on \zh.
    The latter dataset was never used in training to ensure the models are able to generalize across datasets.
    Although \z decays are allowed in the \zh sample, the fraction of such events is $\mathcal{O}(3\%)$ and thus does not affect our conclusions.
    
    Similarly to Ref.~\cite{holmberg2022jet} we use $\log{\left(\ptvistrue/\ptjet\right)}$ as the target for momentum reconstruction, since the logarithm of the ratio of the total visible transverse momentum of the $\tau$ components compared to the jet \ptmomentum is distributed approximately normally.
    We use the Huber loss~\cite{10.1214/aoms/1177703732} for the regression task, as the energy response is asymmetrical, and it is less sensitive to outliers than the mean squared error.
    For the decay mode classification task we one-hot encode $\mathrm{DM}^{\mathrm{true}}$ and use the standard cross-entropy loss for the multi-classification task.
    
    The trainings are performed over a maximum of 100 epochs with a batch size of 1024 using the AdamW~\cite{loshchilov2017decoupled} optimizer.
    We cross-validate the training by redoing it three times on different subsets of the training dataset, and using different neural network initializations.
    Each training runs for approximately 8--24 hours on a single 8GB Nvidia RTX 2070S GPU.
    No hypertuning is performed at this stage, as our goal is not to find the most optimal configuration for this specific dataset, but rather to demonstrate generically that such ML algorithms are suitable for end-to-end $\tau_h$ reconstruction.
    
\begin{figure}[H]
        \centering
        \includegraphics[width=0.48\textwidth]{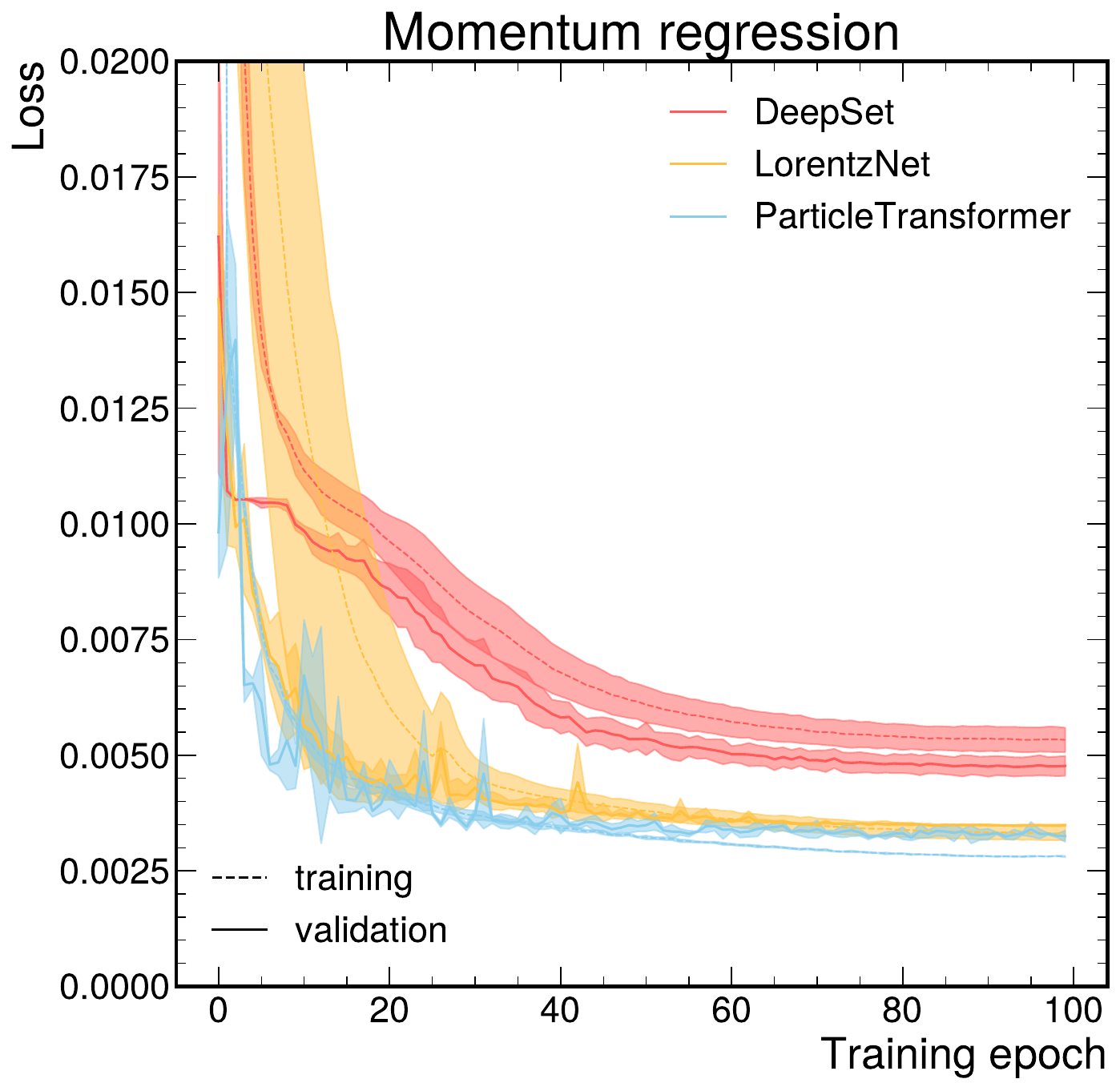}
        \includegraphics[width=0.48\textwidth]{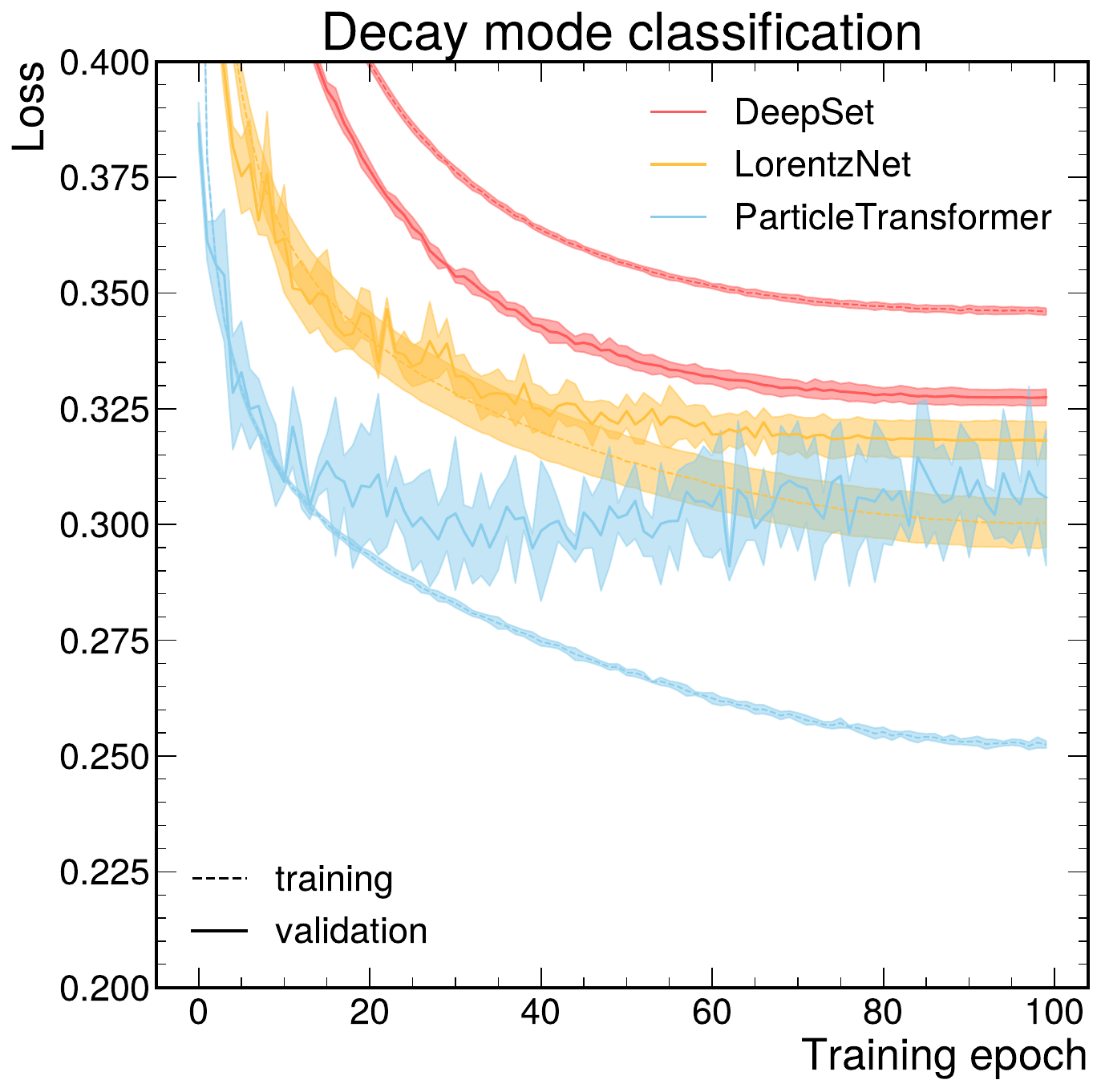}\\
        \includegraphics[width=0.48\textwidth]{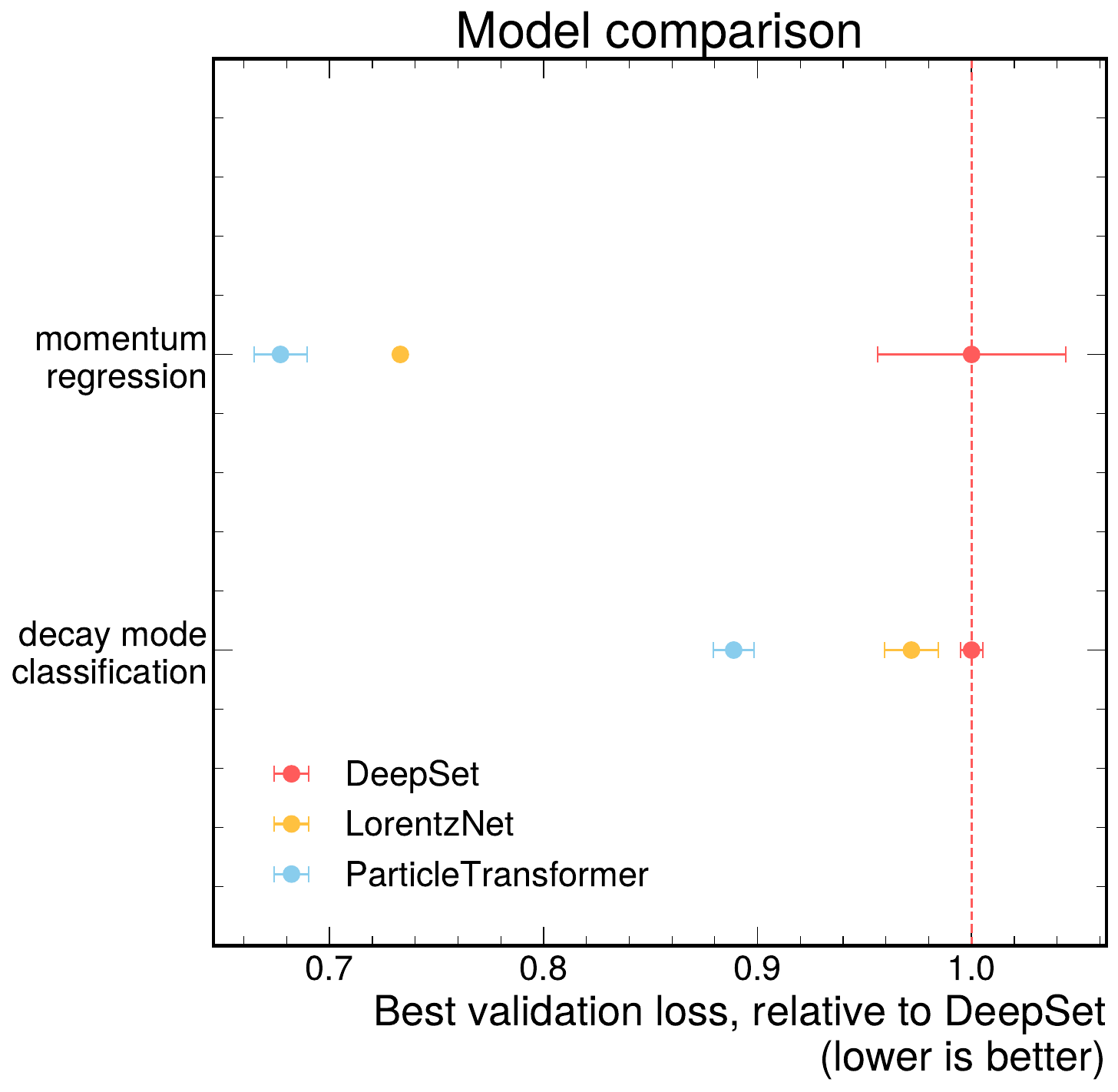}
        \caption{ 
        Training and validation loss curves for the DeepSet, LorentzNet and ParticleTransformer algorithms in the $\tau_h$ momentum regression task (\textbf{Top-Left}) and the $\tau_h$ decay mode classification task (\textbf{Top-Right}). \textbf{Bottom:} Comparison of the best achieved validation losses for the three different algorithms in both the $\tau_h$ momentum regression and $\tau_h$ decay mode classification task.}
        \label{fig:losses}
    \end{figure}

    The top part of Fig.~\ref{fig:losses} shows the loss curves for all three algorithms for both tasks.
    As can be seen, more expressive ParticleTransformer and LorentzNet architectures converge both faster than the simpler DeepSet algorithm and also achieve overall lower ultimate validation losses, as shown in the bottom part of the same figure.
    
\section{Results}\label{sec:results}
\begin{figure}[H]
            \centering
            \includegraphics[width=0.8\textwidth]{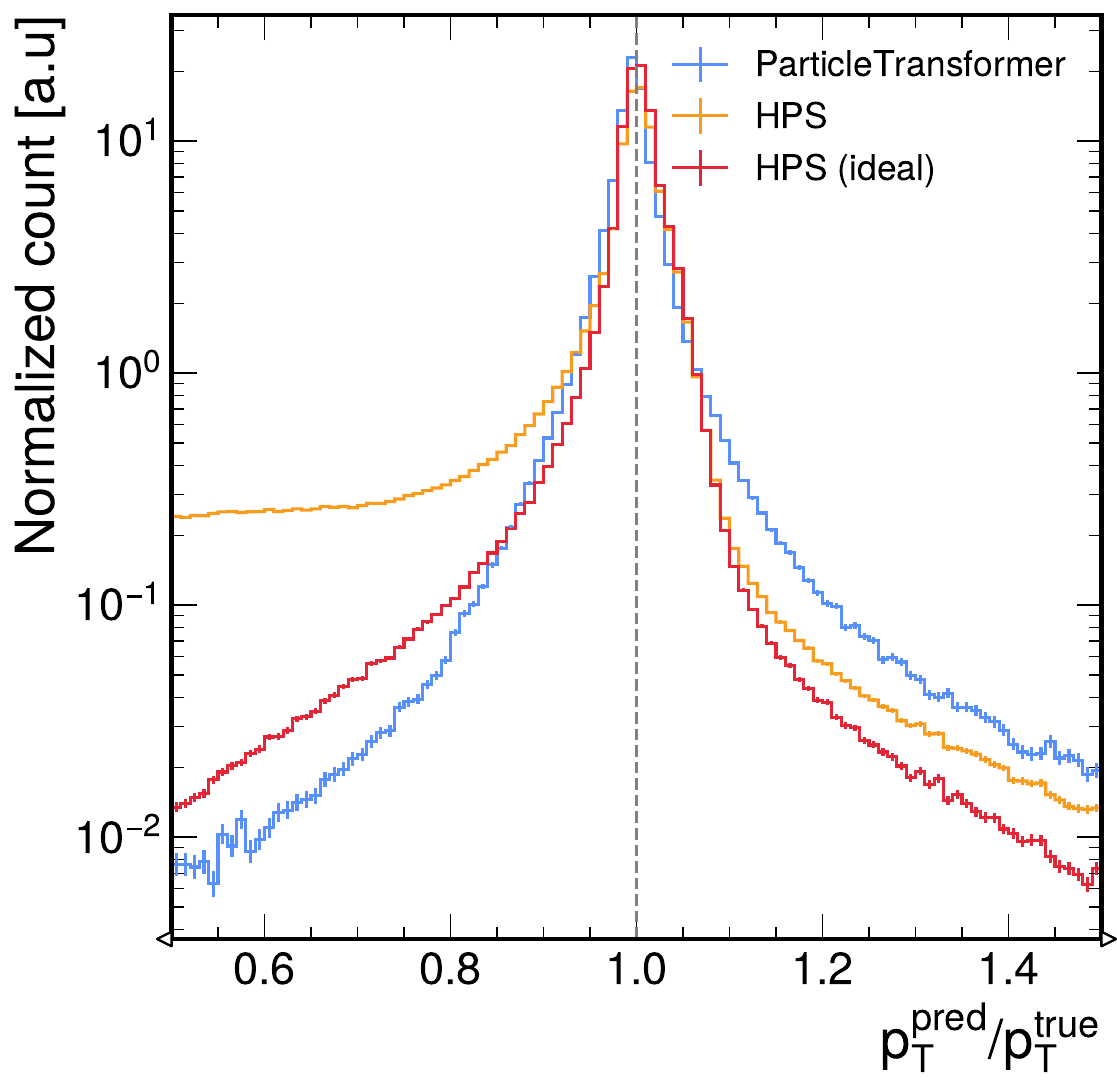}
            \caption{Distribution of the ratio between predicted and true $\tau_h$ transverse momentum for ParticleTransformer and HPS, where HPS tends to underpredict. This underprediction is the result of an incorrect decay mode reconstruction.}
            \label{fig:regression_bin}
        \end{figure}

\begin{figure}[H]
            \centering
            \includegraphics[width=0.48\textwidth]{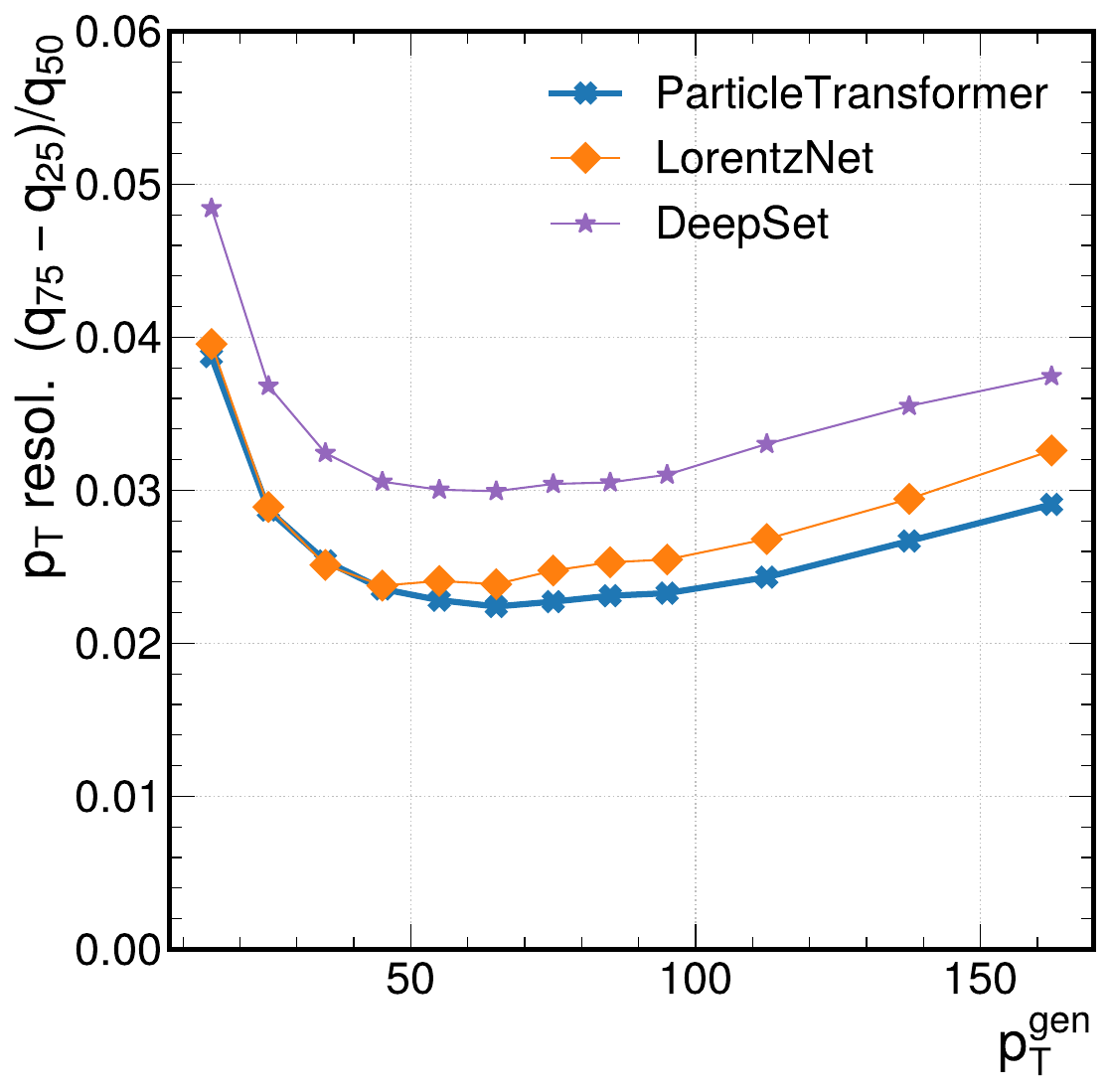}
            \includegraphics[width=0.48\textwidth]{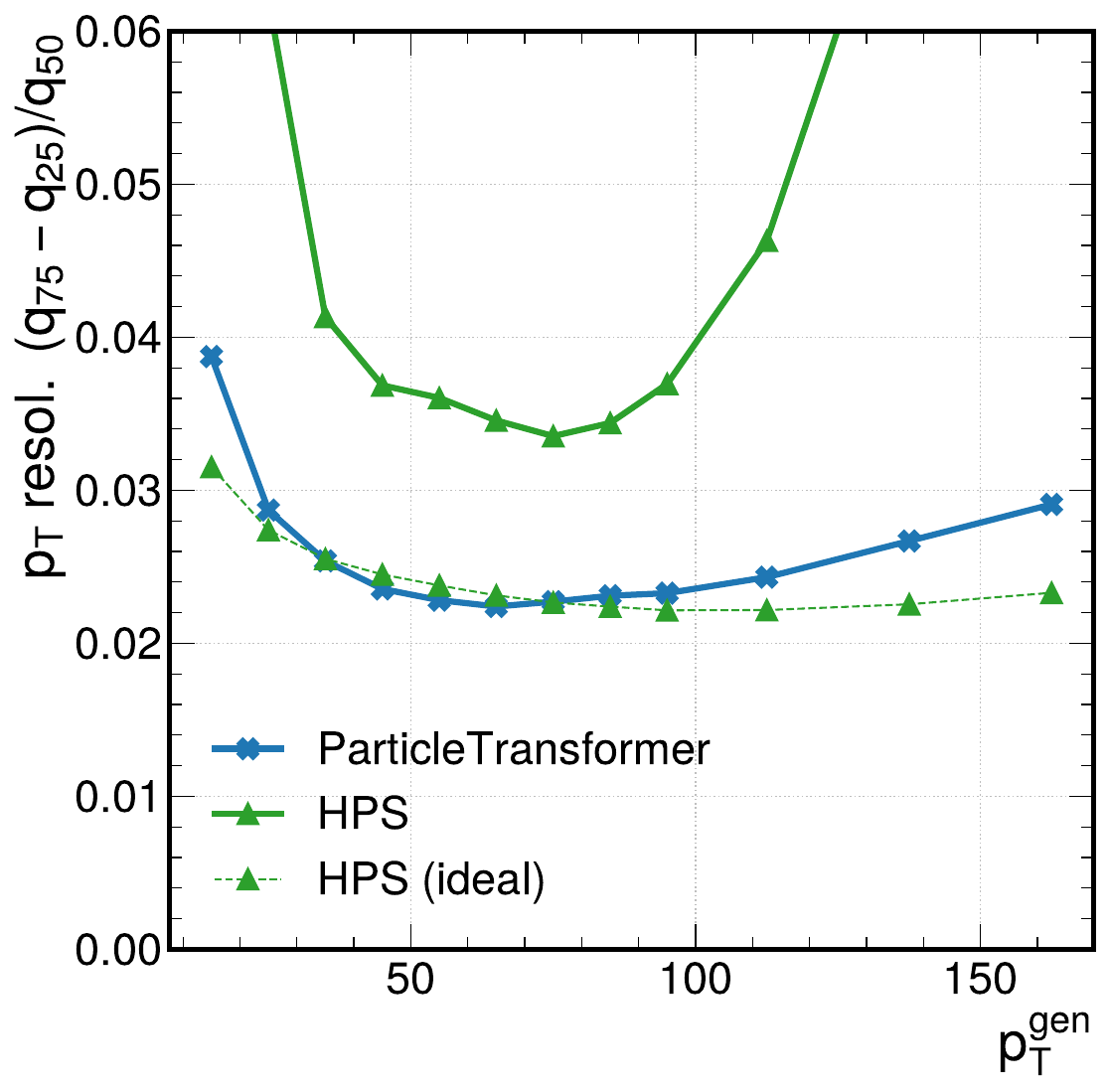}
            \caption{The resolution of the $\tau_h$ momentum response distribution given by the \gls{iqr} of the $\ptvispred/\ptvistrue$ distribution of the tested ML models (\textbf{Left}) and comparison of the best model to the heuristic baseline HPS (\textbf{Right}). The response is as a function of the truth level visible $\tau$ \ptmomentum for the \zh sample for all algorithms.}
            \label{fig:results_regression}
        \end{figure}

    To quantify the performance of the three ML algorithms in the $\tau_h$ momentum regression it is useful to measure the resolution of the resulting $\tau_h$ momentum distribution.
    The resolution is given by the width of the $\ptvispred/\ptvistrue$ distribution.
    For our results we use the \gls{iqr} instead of the standard deviation as a measure for the width as it is less sensitive to outliers.
    The \gls{iqr} is given by the difference in the position of the 25\% and 75\% quantile of the distribution, normalized by the 50\% quantile, giving the width of the central part of the distribution relative to its median.
    Fig.~\ref{fig:regression_bin} compares the predicted and true $\tau_h$ $\ptvispred/\ptvistrue$ ratio distributions for \gls{hps} and ParticleTransformer, where we see that due to the failure of the decay mode prediction, the HPS algorithm has a significant amount of underpredictions compared to the best ML algorithm.
    Fig.~\ref{fig:results_regression} shows the resolution of the regressed \ptvispred distribution as a function of the generator-level $\tau_h$ \ptvistrue separately for all three ML algorithms, and \gls{hps} against the best-performing model, i.e. the ParticleTransformer.

    Similarly, all three ML algorithms achieve a resolution of about 2.1\% to 3\%, which is close to the ideal case of HPS, if the decay mode was correctly reconstructed. The degradation of the resolution of HPS is due to the cases where HPS combinatorially reconstructs the wrong decay mode. We find that the ParticleTransformer algorithm performs the best, and is not affected by the combinatorial problem of HPS.
    This compares to a resolution of about 3.5\% to more than 10\% in the same $p_T$ range when using HPS.
    At high $\tau_h$ transverse momentum  ($\ptvistrue>100$ GeV), where the number of available jets in the \z sample used for training drops significantly, the performance of all three ML algorithms, but in particular the ParticleTransformer and LorentzNet, is reduced, under-predicting the $\tau_h$ \ptvistrue by up to a few percent while the resolution degrades to around 3\%.
    
\begin{figure}[H]
        \centering
        \includegraphics[width=0.9\columnwidth]{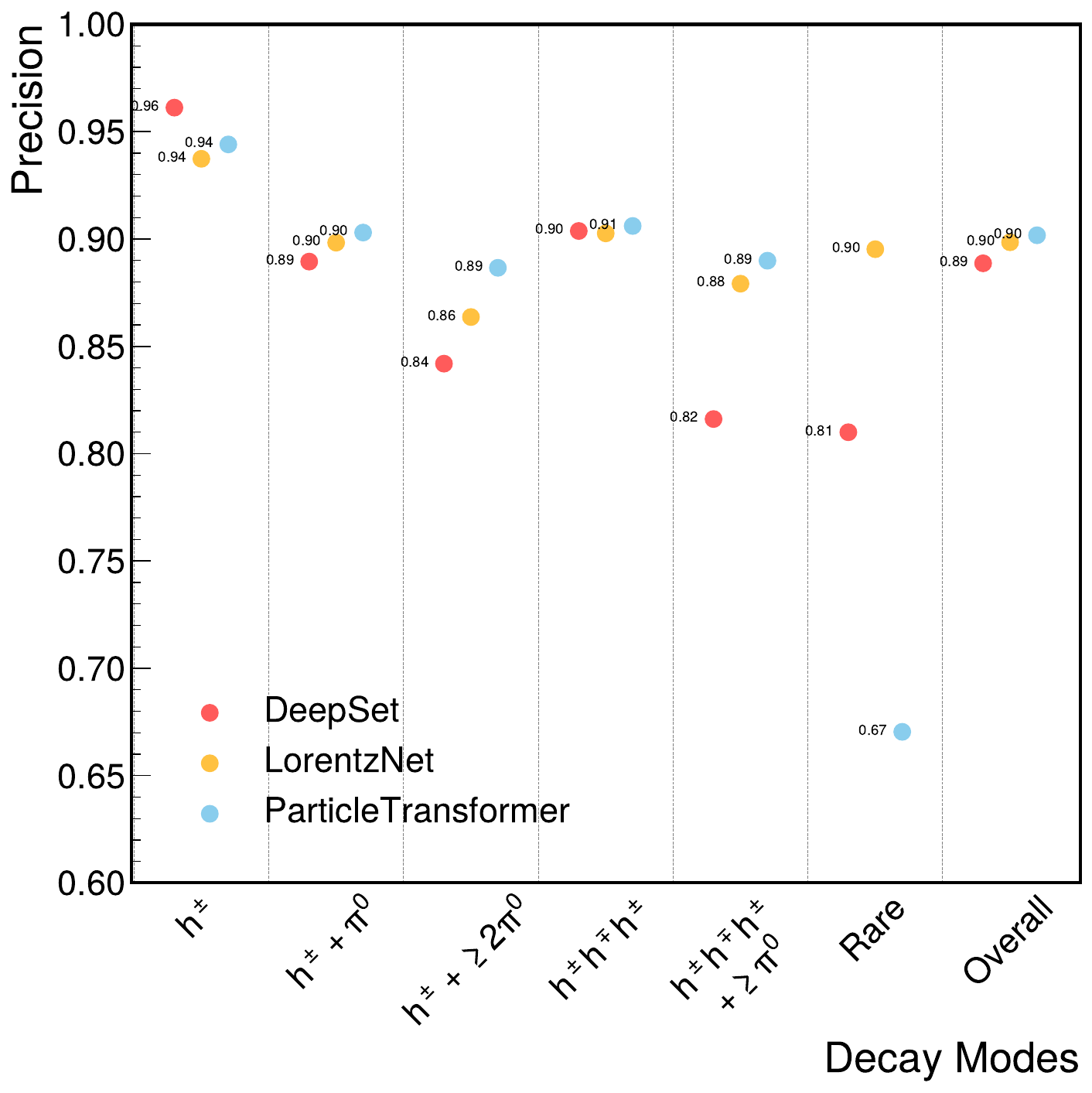}
        \caption{The precision of $\tau_h$ decay mode reconstruction in channels with a varying number of charged and neutral particles in the \zh dataset. The ``Overall'' category represents the general performance of a model across all decay modes weighted by the branching ratio of the given decay mode.}
        \label{fig:results_classification-dm}
    \end{figure}

\begin{figure}[H]
        \centering
        \includegraphics[width=0.8\columnwidth]{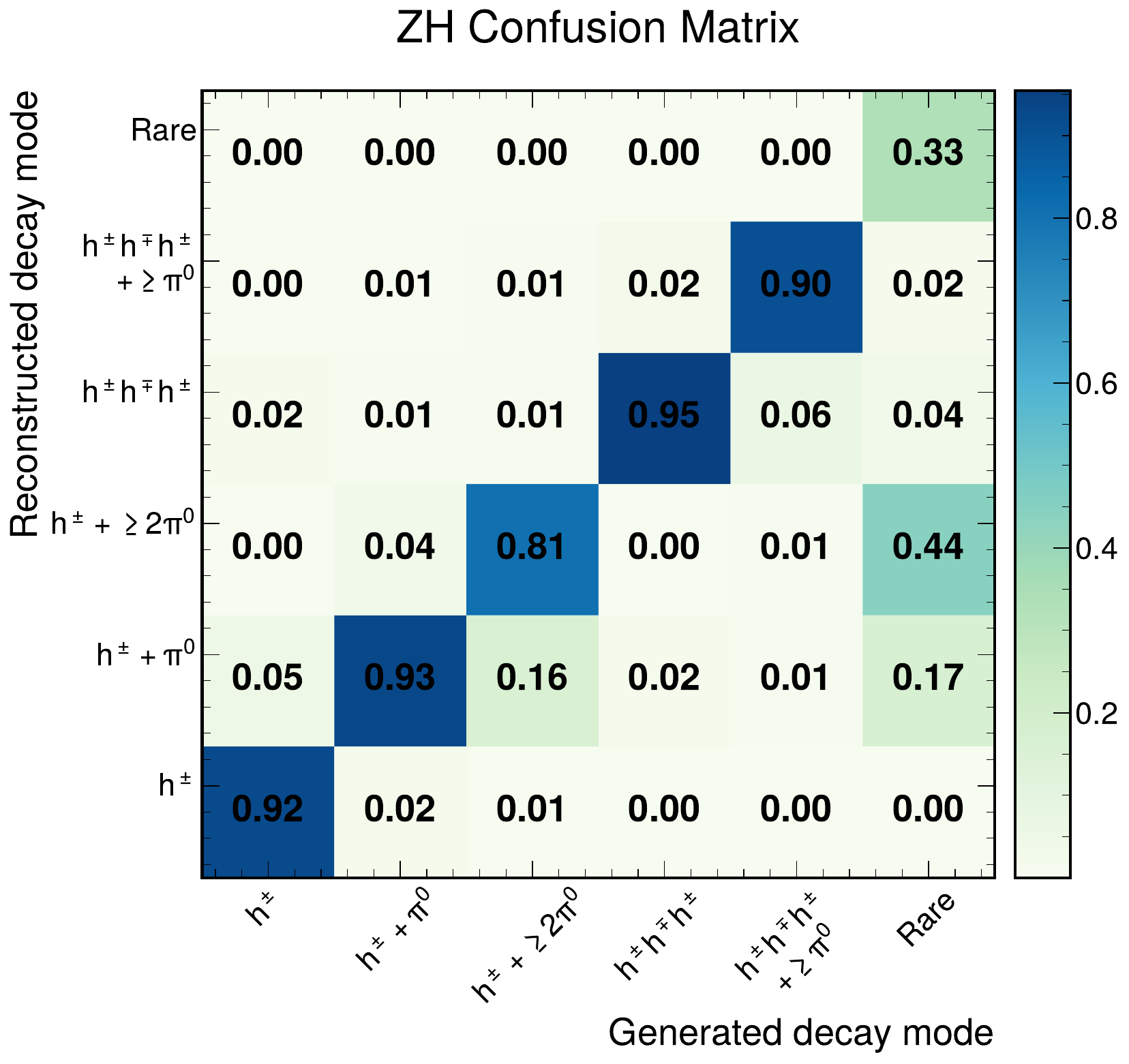}
        \caption{Confusion matrix of $\tau_h$ decay modes for ParticleTransformer on the \zh dataset.}
        \label{fig:results_classification-cm}
    \end{figure}

    Quantifying the quality of the $\tau_h$ decay mode classification is a more difficult task as the fraction of $\tau$ leptons for the different decay modes varies, with the most likely decay into one charged hadron and one neutral pion at about 26\% of all $\tau$ lepton decays, and decays into three charged hadrons and one neutral pion at only about 5\%.
    On the other hand, depending on the underlying physics analysis, the identification of the right number of charged hadrons, and subsequently the correct charge of the $\tau_h$, might be physically more interesting than the differentiation of decay modes with zero, one or two neutral hadrons.
    In Fig.~\ref{fig:results_classification-dm} we show the precision broken down by decay mode for all three algorithms.
    The confusion matrix of true and reconstructed decay modes for the ParticleTransformer algorithm is shown in Fig.~\ref{fig:results_classification-cm}.
    The latter breaks down the precision of the algorithm as a function of the decay mode, and helps us to judge the typical cases of misidentified decay modes.
    We have chosen the ParticleTransformer algorithm as it shows the best overall loss, as can be seen in Fig.~\ref{fig:losses}.

    Overall a precision of 80\% to 95\% is achieved for decay mode classification, with the most difficult decay modes being the ones with multiple neutral pions, as well as rare decays.
    In general, the ParticleTransformer algorithm performs the best for these difficult cases with a high number of final state particles.
    Notably, ParticleTransformer algorithm performs best out of the tested models in all decay modes except the statistically less populated category (``Rare'').
    At the same time, the ParticleTransformer has comparable or better decay mode reconstruction performance than the combinatorial HPS algorithm. 
    We hypothesize that the ParticleTransformer approach is affected more significantly than the other networks by the limited training data.
    Nevertheless, given the broadly comparable overall precision of the models in the decay mode reconstruction task, the choice of a particular architecture should be driven by the available computational budget.
        
\section{Summary and Outlook}\label{sec:outlook}
    In this paper we define $\tau_h$ reconstruction as a multi-task machine learning problem consisting of binary classification for $\tau_h$ identification, momentum regression, and decay mode multi-classification, create a realistic benchmark dataset for this task and compare the performance of several architectures on the reconstruction sub-tasks.
    The $\mathtt{Fu}\tau\mathtt{ure}$ dataset is available in Ref.~\cite{dataset} and contains samples to test machine learning algorithms for $\tau_h$ reconstruction.
    This paper builds on previous work~\cite{Lange:2023gbe} showing how transformer-based architectures can be used to reliably reconstruct and identify hadronic $\tau$ decays.
    Here, we demonstrate that these architectures also perform well in regressing the $\tau_h$ momentum as well as identifying the $\tau_h$ decay mode, thus facilitating the treatment of $\tau_h$ identification and reconstruction as a single machine learning problem.
    For the momentum regression a resolution of about 3\% with a momentum scale about 0.5\%-1\% within the true $\tau_h$ momentum could be achieved for the bulk of the its momentum distribution.
    Depending on the specific decay mode a precision between 80\% and 95\% could be achieved for the classification task, with ParticleTransformer outperforming other architectures for the more difficult decay mode with a higher number of final state particles.
    
    Thus far, we have trained separate models for each sub-task from scratch, while recent work in the direction of foundation models has shown promise that using pre-trained backbone models with task-specific fine-tuning can reduce the amount of required training samples, as well as the required inference budget~\cite{Mikuni:2024qsr,Birk:2024knn,Golling:2024abg}.
    In future work, it may be useful to investigate the dependence of the architectures on available training statistics and sample composition to ensure robustness under domain shift scenarios.
    The published dataset enables studies of trade-offs between bias and variance in terms of using physics-inspired networks, such as LorentzNet, on a limited set of input features, vs. using a wide variety of input features in more generic architectures, such as ParticleTransformer, to identify the relative feature importances for the sub-tasks.
    The dataset and training and validation setup can be generalized in a straightforward way to accommodate FCC-ee and other future collider scenarios.

\section*{Acknowledgments}
    We would like to thank Karl Ehatäht for reviewing the manuscript and providing useful feedback.
    This work has been supported by the Estonian Research Council grants
    PSG864, % Joosep's grant
    RVTT3-KBFI % CERN Eesti teaduse konsortsium - CERN CMS
    and by the European Regional Development Fund through the CoE program grant TK202. % CoE program "Fundamental Universe". 

\section*{Data availability}
    The dataset used in this paper is made available in Ref.~\cite{dataset}. The software used to produce the results can be found in Ref.~\cite{software}.

\section*{Author contributions (CRediT)} % https://www.elsevier.com/researcher/author/policies-and-guidelines/credit-author-statement
    \textbf{Laurits Tani}: Conceptualization, Methodology, Software, Validation, Formal analysis, Investigation, Data Curation, Writing - Original Draft, Writing - Review \& Editing, Visualization.
    \textbf{Nalong-Norman Seeba}: Software, Validation, Formal analysis, Investigation, Visualization.
    \textbf{Joosep Pata}: Conceptualization, Methodology, Software, Validation, Formal analysis, Investigation, Data Curation, Writing - Original Draft, Writing - Review \& Editing, Visualization, Supervision, Project administration, Funding administration.
    \textbf{Torben Lange}: Conceptualization, Methodology, Formal analysis, Writing - Original Draft.
    \textbf{Hardi Vanaveski}: Software, Validation, Formal analysis, Investigation, Visualization.

\printbibliography
\end{document}